\documentclass{PoS}

\makeatletter

\def\compoundrel#1\over#2{\mathpalette\compoundreL{{#1}\over{#2}}}
\def\compoundreL#1#2{\compoundREL#1#2}
\def\compoundREL#1#2\over#3{\mathrel
	{\vcenter{\hbox{$\m@th\buildrel{#1#2}\over{#1#3}$}}}}
\makeatother
\newcommand{\bfi}[1]{\mbox{\boldmath $#1$}}
\usepackage{graphicx}

\title{Lambda-nucleon force from lattice QCD }

\ShortTitle{Lambda-nucleon force from lattice QCD }

\author{\speaker{Hidekatsu Nemura}%
	 \\
	 Strangeness Nuclear Physics Laboratory,
	 Nishina Center for Accelerator-Based Science, RIKEN,%
	 Wako, %
	 351-0198, Japan \\
        E-mail: \email{nemura@riken.jp}}

\author{%
	Noriyoshi Ishii\\
	Center for Computational Sciences,
	University of Tsukuba, %
	Tsukuba 305-8571, Japan \\
	E-mail: \email{ishii@ribf.riken.jp}
	}

\author{%
	Sinya Aoki\\
	Graduate School of Pure and Applied Sciences,
	University of Tsukuba, %
	Tsukuba 305-8571, Japan \\
	Riken BNL Research Center, Brookhaven National Laboratory, %
	Upton, New York 11973, USA \\
	E-mail: \email{saoki@het.ph.tsukuba.ac.jp}
	}

\author{%
	Tetsuo Hatsuda\\
	Department of Physics, University of Tokyo, %
	Tokyo 113-0033, Japan \\
	E-mail: \email{hatsuda@phys.s.u-tokyo.ac.jp}
	}

\author{%
        for PACS-CS Collaboration\\
	}

\abstract{%
We study the $\Lambda$-nucleon ($\Lambda N$) force by using lattice QCD.
The Bethe-Salpeter amplitude is calculated 
for the lowest scattering state of the $\Lambda N$ 
so as to obtain the $\Lambda N$ potential. 
The numerical calculation is twofold:
	  (i) Full lattice QCD by using $2+1$ flavor PACS-CS 
	  gauge configurations with, $\beta=1.9$,
	  corresponding to the lattice spacing of $a=0.0907(13)$ fm,
	  on a $32^3\times 64$ lattice.
	  A set of parameter $(\kappa_{ud},\kappa_s)=(0.13770,0.13640)$
	  is used,
	  which corresponds to
	  $m_\pi\approx 300$ MeV and
	  $m_K\approx 594$ MeV.
	  The spatial lattice volume corresponds to (2.86 fm)$^3$. 
	  (ii) Quenched lattice QCD with $\beta=5.7$,
	  the lattice spacing of $a=0.1416(9)$ fm,
	  on the $32^3\times48$ lattice. 
	  Two sets of hopping parameters
	  $(\kappa_{ud},\kappa_s)=(0.1665,0.1643),(0.1670,0.1643)$
	  are used.
	  The spatial lattice volume is (4.5 fm)$^3$. 
For the full QCD,
we find that the $\Lambda p$ has a relatively strong (weak) repulsive
core in the $^1S_0$ ($^3S_1$) channel at short distance,
while the potential has slight attractive region at medium distance. 
The lowest scattering energy in the finite lattice volume is 
calculated; 
Slightly negative values obtained in both spin channels. 
For the quenched QCD, 
we find that the results are qualitatively in agreement with those 
in the full QCD calculation. 
	  }

\FullConference{The XXVI International Symposium on Lattice Field Theory \\
		 July 14 - 19, 2008\\
		 Williamsburg, Virginia, USA}

\begin{document}

\section{Introduction}

Study of the hyperon-nucleon ($YN$) and hyperon-hyperon ($YY$) 
interactions is one of the important tasks 
in the contemporary nuclear physics. 
These interactions are the bases to 
explore the strange nuclear systems, in which 
hyperons (or strange quarks) are embedded in normal 
nuclear systems as ``impurities''~\cite{Hashimoto:2006aw}. 
Various issues, such as 
spin- and flavor-dependence, 
effect of the flavor $SU(3)$ breaking, 
effect of hyperonic mixing 
(e.g., $\Lambda N-\Sigma N$ and $\Lambda\Lambda-\Xi N$), 
are the topics to be solved experimentally and/or theoretically. 
For example, 
recent systematic study (e.g., Ref.~\cite{Nemura:2002fu}) 
for light ($s$-shell) $\Lambda$ hypernuclei 
($^3_\Lambda$H, $^4_\Lambda$H, $^4_\Lambda$He and 
$^5_\Lambda$He) 
suggests that the 
$\Lambda N$ interaction in the $^1S_0$ channel is more
attractive than that in the $^3S_1$ channel. 
However, 
the present $YN$ and the $YY$ 
interactions 
have large uncertainties, because the scattering experiments 
are either difficult or impossible due to the short life-time of 
hyperons. 
Theoretically, it should be desirable 
to understand the $YN$ and $YY$ interaction 
(or, in more general, baryon-baryon interaction)
based on 
the dynamics of quarks and gluons as 
fundamental degrees of freedom. 
If one can perform such an appropriate deduction 
along the theory of quantum chromodynamics (QCD), 
they should have a reliable prediction regarding 
the $YN$ and $YY$ potentials.

Recently, the lattice QCD studies have been performed for
not only the $NN$ potential but also the $\Xi N$
potential in quenched QCD~\cite{Ishii:2006ec,Nemura0806.1094}. 
See also Refs.~\cite{Aoki:2008hh,AokiProcLat08,IshiiProcLat08} 
for the recent developments. 
The lattice QCD calculations focusing on the scattering parameters 
for $\Lambda N$ system 
based on L\"{u}scher's formula have been reported in 
Refs.~\cite{Muroya:2004fz, BeaneYN2007}. 
The purpose of this report is to calculate the 
$\Lambda N$ potentials 
by using the full and quenched QCD gauge configurations. 
The main results of this work are obtained by using the 
$2+1$ flavor PACS-CS gauge configurations with the
spatial lattice volume $(2.86\mbox{fm})^3$~\cite{PACSCS0807.1661}. 
We also report the results calculated by using quenched lattice QCD 
with larger spatial lattice volume $(4.5\mbox{fm})^3$.

\section{Formulation}

The basic formulation is already given in 
Refs.~\cite{Ishii:2006ec,Nemura0806.1094,Aoki:2008hh}.
(See also Refs.~\cite{Luscher,CPPACS.PRD71_94504_2005}.)
In this report, 
we briefly explain the procedure to obtain the potential 
from lattice QCD together with some technical point newly 
introduced in the present work. 
We start from an effective Schr\"{o}dinger equation for the
equal-time Bethe-Salpeter (BS) wave function:
\begin{equation}
 -{1\over 2\mu}\nabla^2 \phi(\vec{r}) +
  \int U(\vec{r},\vec{r}^\prime)
  \phi(\vec{r}^\prime) d^3r^\prime  =
  E \phi(\vec{r}).
\end{equation}
Here $\mu=m_{\Lambda}m_{N}/(m_{\Lambda}+m_{N})$ and 
$E\equiv k^2/(2\mu)$ are the reduced mass of the $\Lambda N$ system and 
the non-relativistic energy in the center-of-mass frame, respectively. 
We consider the low-energy scattering state so that the 
nonlocal potential can be rewritten 
 by derivative expansion~\cite{TW67},
$
U(\vec{r},\vec{r}^\prime)=
 V_{\Lambda N}(\vec{r},\vec{\nabla})\delta(\vec{r}-\vec{r}^\prime).
$
The general expression of the potential $V_{\Lambda N}$ 
is known to be~\cite{JJdeSwart1971} 
\begin{eqnarray}
 V_{\Lambda N} &=&
  V_0(r)
  +V_\sigma(r)(\vec{\sigma}_{\Lambda}\cdot\vec{\sigma}_{N})
  +V_T(r)S_{12}
  +V_{LS}(r)(\vec{L}\cdot\vec{S}_+)
  +V_{ALS}(r)(\vec{L}\cdot\vec{S}_-)
  +{O}(\nabla^2).
  \label{GenePotNL}
\end{eqnarray}
Here
$S_{12}=3(\vec{\sigma}_{\Lambda}\cdot\vec{n})(\vec{\sigma}_{N}\cdot\vec{n})-\vec{\sigma}_{\Lambda}\cdot\vec{\sigma}_{N}$
is the tensor operator with $\vec{n}=\vec{r}/|\vec{r}|$,
$\vec{S}_{\pm}=(\vec{\sigma}_{N} \pm \vec{\sigma}_{\Lambda})/2$  are %
symmetric ($+$) and antisymmetric ($-$) spin operators,
$\vec{L}=-i\vec{r}\times\vec{\nabla}$ is the orbital %
angular momentum operator. %
We note that the antisymmetric spin-orbit forces 
($V_{ALS}$ and $V_{ALS\tau}$) 
do not arise %
in the $NN$ case because of the identical nature of the 
nucleon within the isospin symmetry.

First we calculate the four-point correlator by using the lattice QCD
\begin{equation}
\label{eq:4-point}
 G_{\alpha\beta;\alpha^\prime\beta^\prime}(\vec{r},t-t_0) = 
 \sum_{\vec{X}}
 \left\langle 0
  \left|
   p_\alpha(\vec{X}+\vec{r},t)
   \Lambda_\beta(\vec{X},t)
   \overline{ \Lambda^\prime}_{\beta^\prime}(t_0)
   \overline{ p^\prime}_{\alpha^\prime}(t_0)
  \right| 0 
 \right\rangle,
 \label{4pt.corr}
\end{equation}
where the summation over $\vec{X}$ is to select the state with 
zero total momentum. 
The $p_{\alpha}(x)$ and $\Lambda_{\beta}(y)$ denote the 
interpolating fields for proton and $\Lambda$ 
\begin{eqnarray}
  p_\alpha(x) &=&
  \varepsilon_{abc} \left(
		     u_a(x) C \gamma_5 d_b(x)
		    \right)
  u_{c\alpha}(x),
  \label{proton}
  \\
 \Lambda_\beta(y) &=&
 \varepsilon_{abc}
 \left\{ %
  (d_a(y)C\gamma_5 s_b(y))u_{c\beta}(y)
  + %
  (s_a(y)C\gamma_5 u_b(y))d_{c\beta}(y)
   -2 %
  (u_a(y)C\gamma_5 d_b(y))s_{c\beta}(y)
 \right\}, 
 \label{Lambda}
\end{eqnarray}
and $\overline{ \Lambda^\prime}_{\beta^\prime}(t_0)$ and 
$\overline{ p^\prime}_{\alpha^\prime}(t_0)$ denote the 
wall source
\begin{eqnarray}
 { p^\prime}_{\alpha^\prime}(t) &=&
  \sum_{\vec{x}_1,\vec{x}_2,\vec{x}_3}
  \varepsilon_{abc} \left(
		     u_a(\vec{x}_1,t) C \gamma_5 d_b(\vec{x}_2,t)
		    \right)
  u_{c\alpha^\prime}(\vec{x}_3,t),\\
 { \Lambda^\prime}_{\beta^\prime}(t) &=&
 \sum_{\vec{y}_1,\vec{y}_2,\vec{y}_3} 
 \varepsilon_{abc}
 \left\{ %
  (d_a(\vec{y}_1,t)C\gamma_5 s_b(\vec{y}_2,t))u_{c\beta^\prime}(\vec{y}_3,t)
  + %
  (s_a(\vec{y}_1,t)C\gamma_5 u_b(\vec{y}_2,t))d_{c\beta^\prime}(\vec{y}_3,t)
  \right. \nonumber \\
 && \left. \qquad \qquad \qquad 
   -2 %
  (u_a(\vec{y}_1,t)C\gamma_5 d_b(\vec{y}_2,t))s_{c\beta^\prime}(\vec{y}_3,t)
 \right\}. 
\end{eqnarray}
The Fast Fourier Transformation (FFT) algorithm is imposed 
into the actual {\tt C++} code to reduce the computational time,
which makes the calculation at all of the spatial points ($32^3$) 
possible. 
(In the previous work for the $\Xi^0 p$ system, we calculate only on the
$x$-, $y$- and $z$-axes and the nearest neighbors 
in the long distance region.~\cite{Nemura0806.1094})

In the Monte Carlo calculations, 
noise reductions are made for the four-point correlator 
(\ref{4pt.corr}) obtained from lattice QCD in order to restore 
(i) the rotational (cubic group) symmetry, 
(ii) the spatial reflection symmetry,
(iii) the charge conjugation and time-reversal symmetry.
Then, 
we perform the spin projection for the source 
to obtain the wave function with particular spin component ($J=0,1$). 
\begin{equation}
 F_{p_{\alpha}\Lambda_{\beta}}(\vec{r},t-t_0;J,M)=
  \sum_{\alpha^\prime\beta^\prime}
  P_{\alpha^\prime\beta^\prime}^{(JM)}
  G_{\alpha\beta;\alpha^\prime\beta^\prime}(\vec{r},t-t_0), 
\end{equation}
See Ref.~\cite{IshiiProcLat08} for detail. 
Finally, the desirable wave function $\phi_{\alpha\beta}$
is obtained from the projected four-point correlator
at large $t-t_0$:
\begin{eqnarray}
 F_{p_{\alpha}\Lambda_{\beta}}(\vec{r},t-t_0;J,M) &=&
  \sum_{n} A_{n}
  \sum_{\vec{X}}
  \left\langle 0
   \left|
    p_\alpha(\vec{X}+\vec{r},t)
    \Lambda_\beta(\vec{X},t)
   \right| E_{n} 
  \right\rangle
  e^{-E_{n}(t-t_0)}
  \\
 &\approx&
 A_0 \phi_{\alpha\beta}(\vec{r};J,M) e^{-E_0 (t-t_0)}
 \qquad
 (t-t_0 \rightarrow \mbox{large}). 
\end{eqnarray}
Here $E_n$ ($|E_n\rangle$) is the eigen-energy (eigen-state)
of the six-quark system 
with the particular quantum number $(J^\pi,M)$, 
and 
$A_n = \sum_{\alpha^\prime\beta^\prime} P_{\alpha^\prime\beta^\prime}^{(JM)}
\langle E_n | \overline{ \Lambda^\prime}_{\beta^\prime}
\overline{ p^\prime}_{\alpha^\prime} | 0 \rangle$. 
For simplicity, we consider the effective central potential, 
$V_C(r;J=0)=V_0(r)-3V_\sigma(r)$, for $^1S_0$ 
and, 
$V_C(r;J=1)=V_0(r) +V_\sigma(r)+\cdots$, for $^3S_1$, 
where $\cdots$ is the higher order contribution from 
noncentral force such as $V_T(r)$, 
which is expected to be small in the present calculation. 
The effective central potential $V_C(r;J)$ 
is obtained 
by focusing only on the $S$-wave component of the wave function 
in each spin channel $J$: 
\begin{equation}
 V_C(r;J) = E_J + {1\over 2\mu} {\vec{\nabla}^2\phi(r;J)\over \phi(r;J)}. 
\end{equation}

\section{Numerical calculation}

\subsection{$N_f=2+1$ QCD}

Main results in this report are obtained by using 
the gauge configuration generated by PACS-CS collaboration with 
$2+1$ flavor full QCD~\cite{PACSCS0807.1661}; 
The Iwasaki gauge action and the nonperturbatively $O(a)$-improved 
Wilson quark action are employed. 
Calculations are carried out at $\beta=1.9$ 
on a $32^3\times 64$ lattice, 
corresponding to $a=0.0907(13)$ fm 
where $a$ is the lattice spacing 
at the physical point~\cite{PACSCS0807.1661}. 
A set of hopping parameters 
$(\kappa_{ud},\kappa_s)=(0.13770, 0.13640)$
is used in this work.
Several light hadron masses obtained by PACS-CS 
are shown in Table~\ref{masses}.
The BS wave function is obtained by 
employing the wall source at the time-slice $t_0=0$, 
and the Dirichlet boundary condition is imposed 
along the temporal direction
on the time-slice $t=32$ 
with the Coulomb gauge fixing. 
The present result of the BS wave function is 
calculated with $N_{\rm conf}=422$, 
where $N_{\rm conf}$ is the number of gauge configurations.

\subsection{Quenched QCD with larger spatial volume
  }

In this case, 
we use the plaquette gauge action and the
Wilson fermion action with the gauge coupling
$\beta=5.7$ on the $32^3\times 48$ lattice.
The periodic boundary condition is imposed for
quarks in the spatial direction. 
The wall source is placed at $t_0=0$ 
with the Coulomb gauge fixing and the 
Dirichlet boundary condition is imposed at $t=24$ in the temporal direction. 
These setup are essentially the same as our previous  calculations  
for the $NN$ and $\Xi N$ potentials~\cite{Ishii:2006ec,Nemura0806.1094} 
except for the treatment of the temporal part. 
The lattice spacing 
at the physical point 
is determined as 
$a=0.1416(9)$~fm ($1/a = 1.393(9)$ GeV)
from $m_\rho=770$ MeV. 
The hopping parameter for the strange quark mass 
is given by $\kappa_s=0.16432(6)$ from 
$m_K=494$ MeV. 
The spatial lattice volume is $(4.5\mbox{fm})^3$, 
which is enough to accommodate two baryons. 
The present result of the BS wave function 
in quenched QCD 
is obtained with $N_{\rm conf}=550$.

\section{Results and Discussion}

\begin{table}[b]
 \centering \leavevmode 
 \begin{tabular}{ccccccccc}
  \hline \hline
  & 
  $m_\pi$ & $m_\rho$ & $m_K$ & $m_{K^\ast}$ & 
  $m_N$ & $m_\Lambda$ & $m_{\Sigma}$ & $m_{\Xi}$ \\
  \hline
  \multicolumn{9}{l}{\bf 2+1 flavor QCD by PACS-CS with
  ${\bf (\kappa_{\it ud},\kappa_{\it s})=(0.13770,0.13640)}$
  } \\
  Ref.\cite{PACSCS0807.1661} &
  296(3) & 
  848(20) &
  594(2) &
  985(8) &
  1093(19) &
  1254(14) &
  1315(15) &
  1448(10) \\
  \hline
  \multicolumn{9}{l}{\bf quenched QCD with
  ${\bf \beta=5.7, \kappa_{\it s}=0.1643}$,
  } \\
  $\kappa_{ud}=0.1665$ &
  514(1)  & 859(3)  & 607(1)  & 904(3) &
  1287(8) & 1345(6) & 1368(6) & 1412(5) \\
  $\kappa_{ud}=0.1670$ &
  465(1)   & 840(6)  & 588(1)  & 894(3) &
  1234(11) & 1308(7) & 1341(8) & 1396(5) \\
  \hline
  Exp. & 
  135   &  770   & 494  & 892   &
  940   & 1116   & 1190 & 1320  \\
 \hline \hline 
  \end{tabular}
 \caption{
 Hadron masses in the unit of MeV. 
 The results for $2+1$ flavor QCD by PACS-CS is taken from
 Ref.~\cite{PACSCS0807.1661}.
 $N_{\rm conf}=560$ in quenched QCD. 
 }
 \label{masses}
\end{table}

\subsection{$N_f=2+1$ QCD}

\begin{figure}[t]
 \centering \leavevmode
 \begin{minipage}[t]{0.49\textwidth}
  \includegraphics[width=.9\textwidth]
  {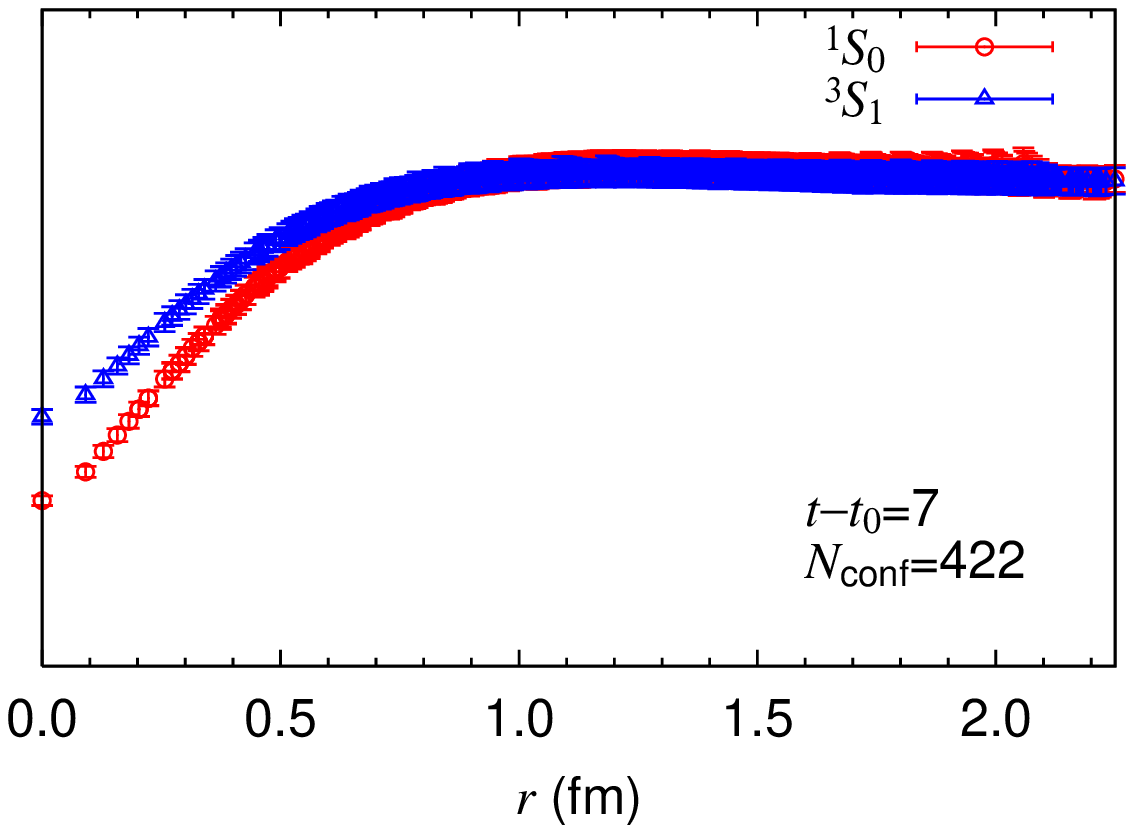}
  \footnotesize
 \end{minipage}
 \hfill
 \begin{minipage}[t]{0.49\textwidth}
  \includegraphics[width=.9\textwidth]
  {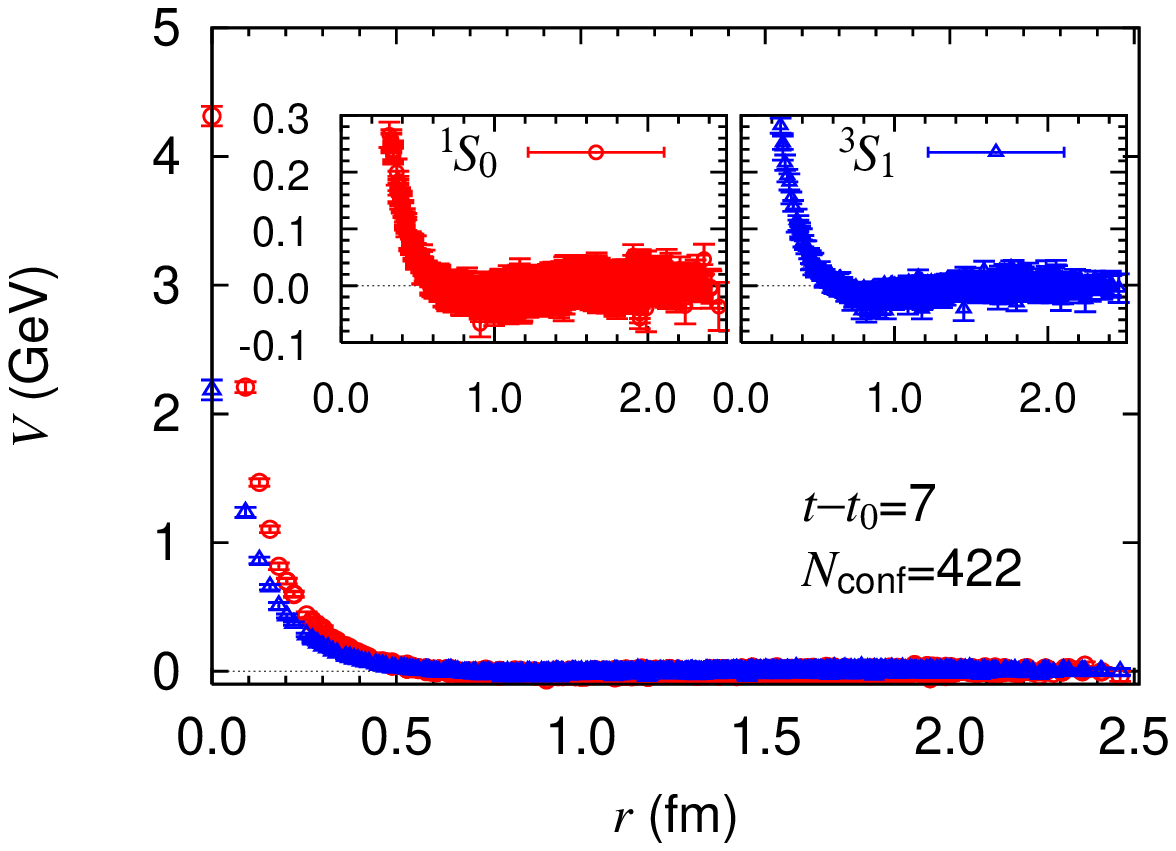}
 \end{minipage}
 \caption{
 Left:
 The radial wave function of $\Lambda p$,
 in $^1S_0$ (red circle) and $^3S_1$ (blue triangle) channels,
 obtained by using the PACSCS gauge configurations at $t-t_0=7$.
 $N_{\rm conf}=422$ gauge configurations are used.
 Right:
 The effective central potential for $\Lambda p$,
 in the $^1S_0$ (red circle) and $^3S_1$ (blue triangle),
 obtained from the wave function at time slice $t-t_0=7$. 
 The left (right) inset shows its enlargement 
 in the $^1S_0$ ($^3S_1$) channel. 
 }
 \label{wavepot_PACSCS}
\end{figure}
\begin{figure}[b]
 \centering \leavevmode
 \begin{minipage}[t]{0.49\textwidth}
  \includegraphics[width=.99\textwidth,angle=0]
  {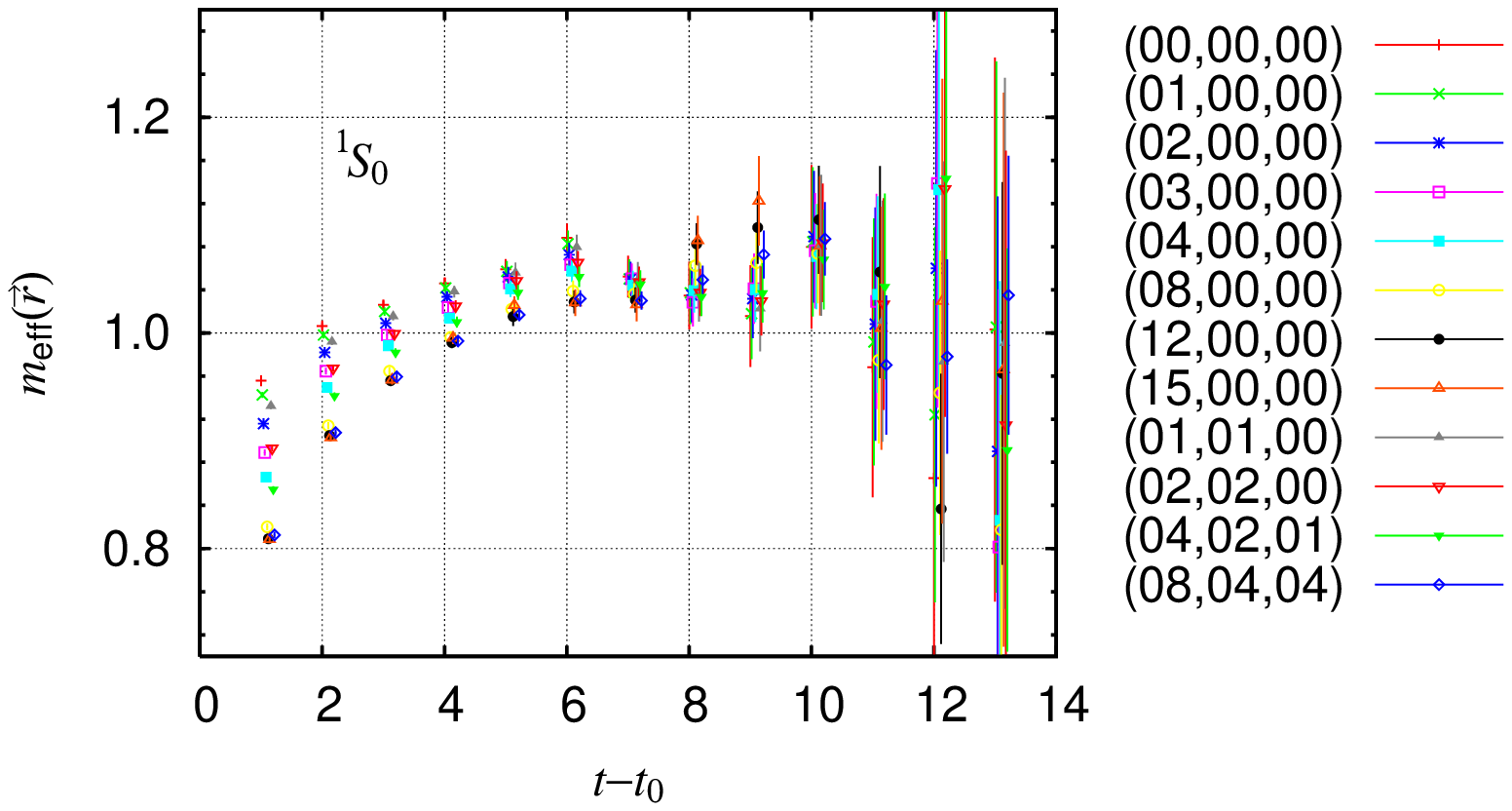}
  \footnotesize
 \end{minipage}
 \hfill
 \begin{minipage}[t]{0.49\textwidth}
  \includegraphics[width=.99\textwidth,angle=0]
  {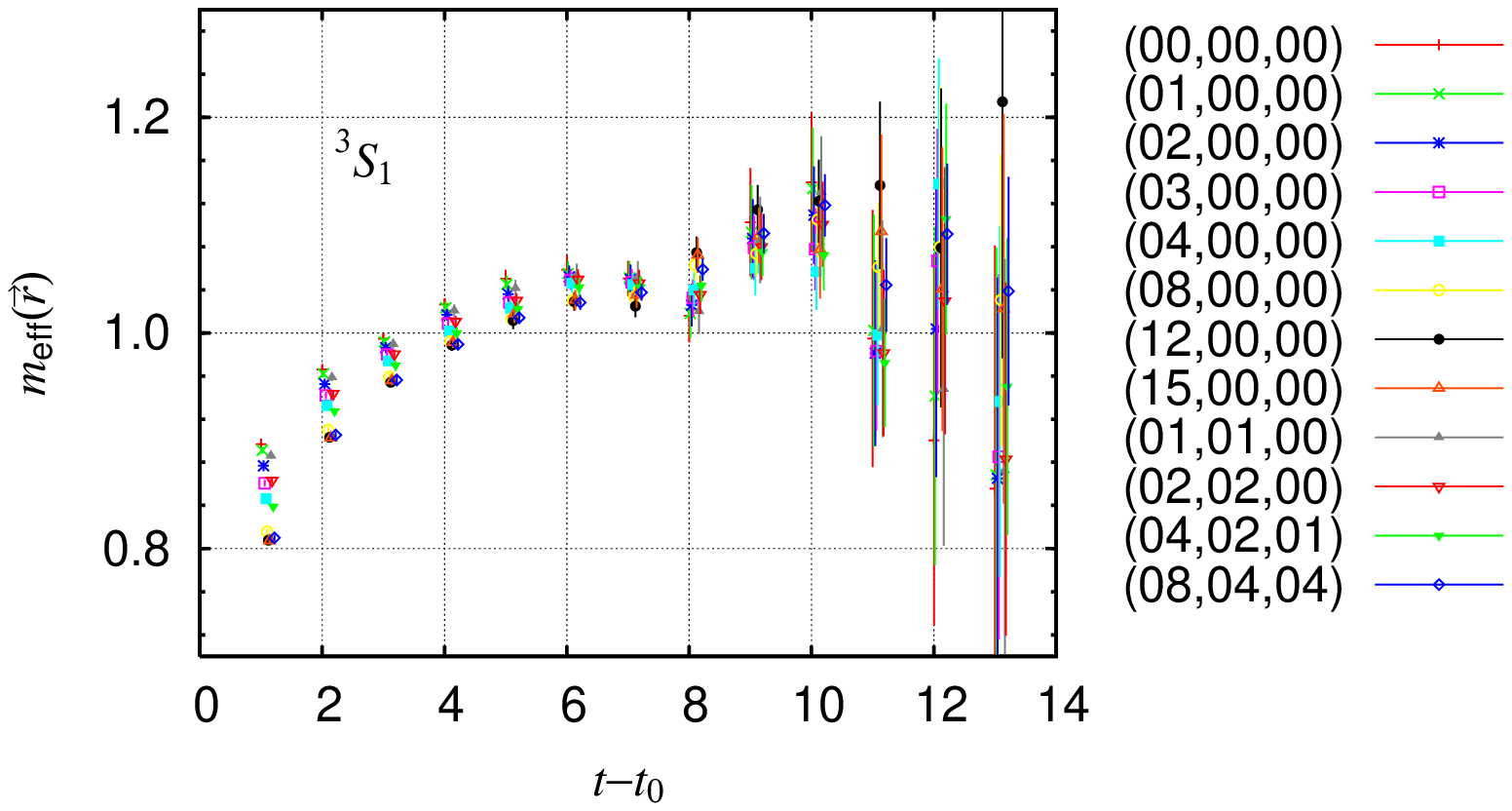}
 \end{minipage}
 \caption{The effective mass of the wave function
 for $\Lambda p$ at several spatial points ($x,y,z$),
 in the $^1S_0$ (left) and the $^3S_1$ (right) channels,
 obtained by using the PACS-CS gauge configurations. 
 $N_{\rm conf}=422$ is used. 
 }
 \label{effmass_PACSCS}
\end{figure}

The left panel of Figure~\ref{wavepot_PACSCS} shows the wave function 
obtained at the time slice $t-t_0=7$. 
The red circle (blue triangle) corresponds to the $^1S_0$ ($^3S_1$)
channel. 
In order to find the ground state saturation of the 
$\Lambda p$ system, 
we 
define 
the ``effective mass'' of the wave function: 
\begin{equation}
 m_{\rm eff}(t-t_0, \vec{r})
  \equiv 
  \log\left(
       {F_{p\Lambda}(\vec{r},t-t_0) 
       \over
       F_{p\Lambda}(\vec{r},(t+1)-t_0)}
      \right).
  \label{effmass_eq}
\end{equation}
Figure~\ref{effmass_PACSCS} shows the effective mass 
at several spatial points. 
The plateaux starting appears at the time slice $t-t_0=6$ 
in the $^3S_1$ channel 
(in the right panel in Fig.~\ref{effmass_PACSCS}) while 
the plateaux starting seems to appear at the time slice around $t-t_0=7$
in the $^1S_0$ channel (in the left panel). 
We need more statistics. 
In this report, we show the results obtained 
at the time slice $t-t_0=7$.

As is shown in Ref.~\cite{CPPACS.PRD71_94504_2005}, 
the non-relativistic energy $E=k^2/(2\mu)$ is determined by 
fitting the wave function in the asymptotic region in terms of 
the Green's function
\begin{equation}
 G(\vec{r}, k^2) =
  {1\over L^3}
  \sum_{\vec{p}\in\Gamma}
  {1\over p^2 - k^2}
  {\rm e}^{i \vec{p}\cdot\vec{r}}, 
  \qquad
  \Gamma = \left\{
	    \vec{p};  \ \vec{p} = \vec{n}{2\pi\over L},
	    \vec{n}\in{\bfi{Z}^3}
	   \right\} ,
  \label{eq:Green}
\end{equation}
which is the solution of 
$(\bigtriangleup + k^2)G(\vec{r}, k^2)=-\delta_L(\vec{r})$
 with $\delta_L(\vec{r})$ being the periodic delta function 
 \cite{Luscher,CPPACS.PRD71_94504_2005}. 
We attempted to make a fit in the asymptotic region 
($15\le |\vec{r}| \le 16$)
at the time slice $t-t_0=7$. 
The fitting region is chosen so that the potential becomes zero 
within the errorbars. 
The energy obtained in the $^1S_0$ ($^3S_1$) channel is 
$E=-1.9(1.3)$ MeV ($E=-1.3(1.2)$ MeV). 
The present results for the energies suggest that the 
spin dependence of the interaction in the low-energy region 
seems to be weak, 
although we need more statistics 
(and perhaps larger spatial volume) 
to make definite conclusion. 
{\bf 
}
The right panel of Figure~\ref{wavepot_PACSCS} shows 
the effective central potential 
of $\Lambda p$ system obtained at the time slice $t-t_0=7$. 
The strong repulsive core is obtained in the $^1S_0$ channel, 
while the repulsive core in the $^3S_1$ channel is relatively weak. 
The left (right) inset shows the enlargement 
in the $^1S_0$ ($^3S_1$) channel; 
A slight attractive well can be seen in the both spin channels.

\subsection{Quenched QCD}

\begin{figure}[t]
 \centering \leavevmode
 \begin{minipage}[t]{0.49\textwidth}
  \includegraphics[width=.9\textwidth]
  {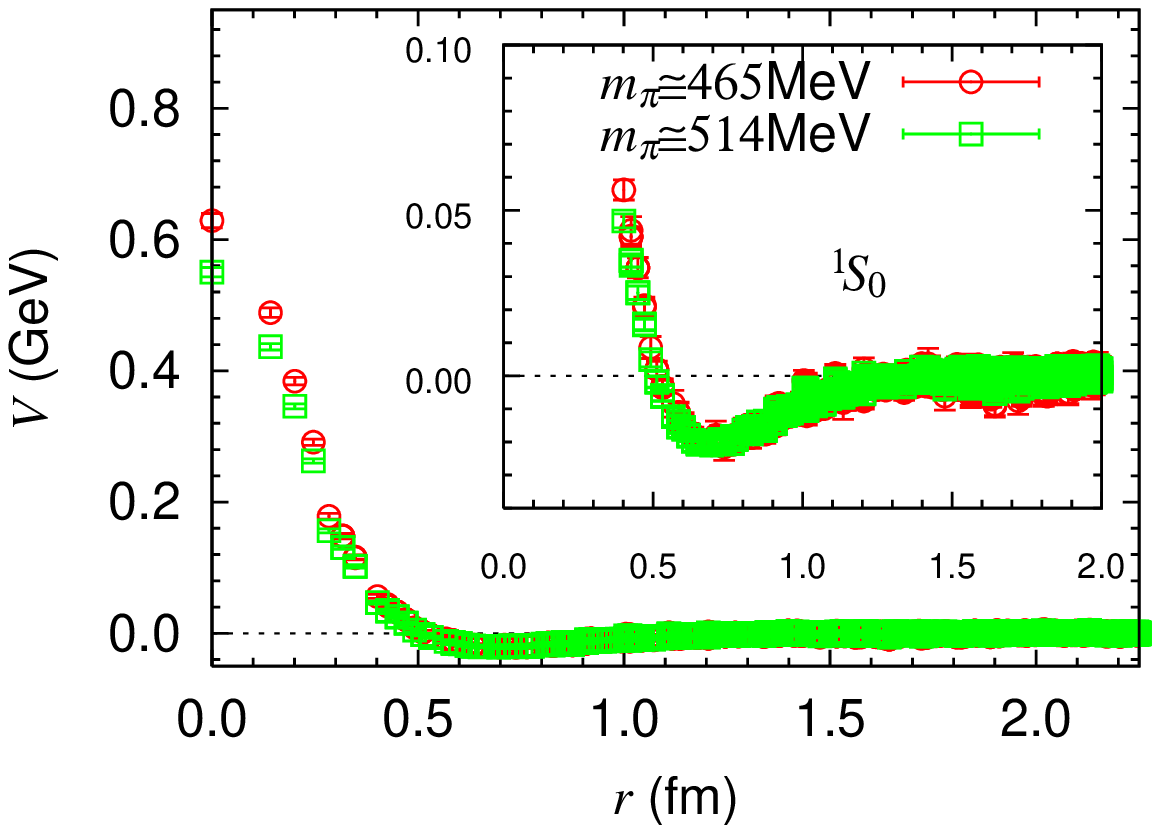}
  \footnotesize
 \end{minipage}
 \hfill
 \begin{minipage}[t]{0.49\textwidth}
  \includegraphics[width=.9\textwidth]
  {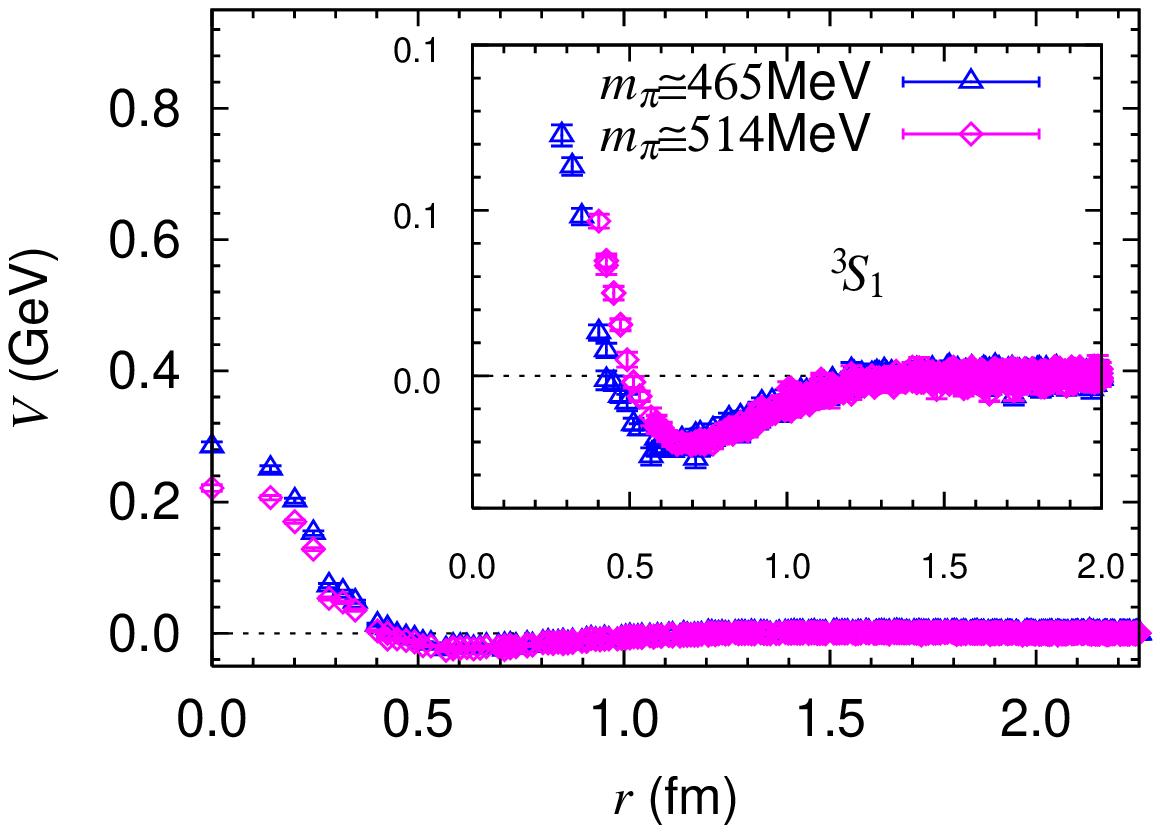}
 \end{minipage}
 \caption{
 Left:
 The effective central potential for $\Lambda p$ in the $^1S_0$ channel 
 at $m_{\pi}\simeq 465$ MeV (red circle) 
 and $m_{\pi}\simeq 514$ MeV (green box).
 Right:
 The effective central potential for $\Lambda p$ in the $^3S_1$ channel
 at $m_{\pi}\simeq 465$ MeV (blue triangle) 
 and $m_{\pi}\simeq 514$ MeV (magenta diamond).
 The inset shows its enlargement. 
 }
 \label{pot_qmassdep}
\end{figure}

We calculate the wave functions for two sets of the 
hopping parameters 
$(\kappa_{ud},\kappa_s)=(0.1665, 0.1643)$ and $(0.1670, 0.1643)$.
Table~\ref{masses} also lists the hadron masses obtained from 
these quenched QCD calculations. 
\begin{figure}[b]
 \centering \leavevmode
  \includegraphics[width=.42\textwidth]
  {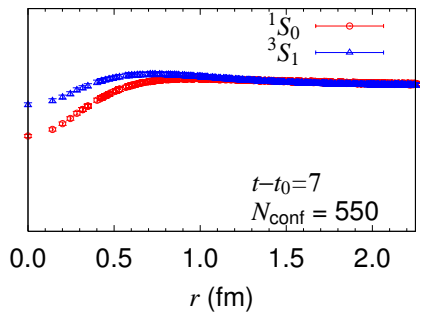}
 \caption{
 Left:
 The radial wave function of $\Lambda p$, in $^1S_0$ (red circle) and
 $^3S_1$ (blue triangle) channels, obtained by quenched QCD 
 calculation with $(\kappa_{ud},\kappa_s)=(0.1670, 0.1643)$, at
 the time slice $t-t_0=7$. 
 }
 \label{wave_ud0.1670s0.1643}
\end{figure}
The Figure~\ref{wave_ud0.1670s0.1643} 
shows the wave function obtained with
$(\kappa_{ud},\kappa_s)=(0.1670,0.1643)$ 
at the time slice $t-t_0=7$,
by using $N_{\rm conf}=550$ gauge configurations. 
For both cases of calculations with 
$(\kappa_{ud},\kappa_s)=(0.1670,0.1643)$ and $(0.1665,0.1643)$ 
in the $^1S_0$ and in the $^3S_1$ channels, 
the ground state saturations are achieved at the time slice 
$t-t_0=7$. 
The energies obtained by fitting the wave function in 
the asymptotic region ($10\le |\vec{r}| \le 16$) 
to the 
Green's function (\ref{eq:Green}) in the $^1S_0$ ($^3S_1$) channel are 
$E=-0.37(12)$MeV ($E=-0.47(11)$MeV) for
($\kappa_{ud},\kappa_s$)=(0.1670,0.1643), 
and 
$E=-0.31(9)$MeV ($E=-0.38(7)$MeV) for
($\kappa_{ud},\kappa_s$)=(0.1665,0.1643),
respectively. 
The lowest scattering energies in both $^1S_0$ and $^3S_1$ channels 
become lower as the $u,d$ quark mass decreases. 
We hardly find the clear conclusion whether the interaction 
in the $^1S_0$ channel 
is more attractive than that in the $^3S_1$ channel. 
Figure~\ref{pot_qmassdep} shows 
the effective central potentials for $\Lambda p$ 
in the $^1S_0$ ($^3S_1$) channel in the left (right) panel,
obtained from the wave function. 
For both ($^1S_0$ and $^3S_1$) spin channels, 
the height of the repulsive core increases as the $u,d$ quark mass 
decreases. 

\bigskip 

In summary, we calculate the $\Lambda N$ potential from lattice QCD.
The full lattice QCD calculation for the $\Lambda N$ system 
with larger volume and smaller lattice spacing 
at physical quark mass 
is highly 
desirable to clarify the spin dependence of the interaction.

\acknowledgments

The authors would 
like to thank  PACS-CS Collaboration for 
allowing us to access their full QCD gauge configurations, 
and 
Dr.~T.~Izubuchi for providing a sample FFT code.
The full QCD calculations have been done 
by using PACS-CS computer 
under the ``Interdisciplinary Computational Science Program'' 
of Center for Computational Science, University of Tsukuba 
(No 08a-12).
The Quenched QCD calculations have been done 
by using Blue Gene/L computer 
under the ``Large scale simulation program''
at KEK (No. 08-19).
Part of numerical analysis has been done by using 
RIKEN super combined cluster system at RIKEN. 
This research was partly supported 
by the Ministry of Education, 
Science, Sports and Culture, Grant-in-Aid 
(Nos. 
18540253, 19540261, 20028013, 20340047).

\end{document}